# Spatio-spectral optical fission in time-varying subwavelength layers

Wallace Jaffray[1], Sven Stengel[1], Fabio Biancalana[1], Colton Bradley Fruhling[2], Mustafa Ozlu[2], Michael Scalora[3], Alexandra Boltasseva[2], Vladimir M. Shalaev[2] and Marcello Ferrera[1]

[1]*Institute of Photonics and Quantum Sciences, Heriot-Watt University, SUPA, Edinburgh, EH14 4AS, UK*
[2] *Elmore Family School of Electrical & Computer Engineering and Birck Nanotechnology Center, Purdue University, West Lafayette, Indiana 47907, USA*
[3]*FCDD-AMT-MGR, DEVCOM AvMC, Charles M. Bowden Research Center, Redstone Arsenal, AL 35898-5000, USA*

## Abstract

Transparent conducting oxides are highly-doped semiconductors that exhibit favourable characteristics when compared to metals, including reduced material losses, tuneable electronic and optical properties, and enhanced damage thresholds. Recently, the photonic community has renewed its attention towards these materials, recognizing their remarkable nonlinear optical properties in the near-infrared spectrum—a feature previously overlooked despite their long-standing application in photovoltaics and touchscreens. The exceptionally large and ultra-fast change of the refractive index, which can be optically induced in these compounds, extends beyond the boundaries of conventional perturbative analysis and makes this class of materials the closest approximation to a time-varying system, and a unique playground for studying a variety of novel phenomena within the domain of photon acceleration. Here we report the spatio-spectral fission of an ultra-fast pulse trespassing a thin film of aluminium zinc oxide with a non-stationary refractive index. By applying phase conservation to this time-varying layer, our model can account for both space and time refraction and explain in quantitative terms, the spatial separation of both the spectrum and energy. Our findings represent an example of extreme nonlinear phenomena on subwavelength propagation distances and shed light on the nature of several nonlinear effects recently reported not accounting for the full optical field distribution. Our work also provides new important insights into transparent conducting oxides' transient optical properties which are critical for the ongoing research in photonic time crystals, on-chip generation of nonclassical states of light, integrated optical neural networks as well as ultra-fast beam steering and frequency division multiplexing.

## 1-Introduction

The spatial engineering of the macroscopic refractive index is broadly exploited in integrated photonics for a myriad of applications including metamaterials, photonic crystals, optical gratings, etc. while temporal engineering remains somewhat underexplored. This is primarily due to two fundamental reasons. First, it is challenging to identify a physical system which can substantially alter its optical properties on ultra-fast time scales. Secondly, time-invariant elements are conceptually more intuitive. Indeed, the modulation of optical signals via electro-optic and all-optical means is broadly used in signal processing. However, the focus of these applications is only on the attained ON/OFF states rather than on the nature of light-matter interaction during the transition time.

The theory of wave propagation in non-stationary media developed starting from the concept of time-varying permittivity [1]. Experimental demonstrations of the effects of time-varying media on light were then introduced in plasma physics under the alias of photon acceleration. Within these settings, noticeable effects can be recorded as photons interact with a plasma undergoing an abrupt electron density transition [2], [3].

In general, when a propagating wave encounters a sudden change of the environmental refractive index, a phenomenon known as time refraction occurs [2], [3], [4], [5]. This effect is analogous to what happens to a wave at a material interface in spatial dimensions, and it shifts photon frequency to satisfy energy conservation as opposed to the spatial case where, the wavevector changes to preserve the momentum (further details in theoretical section). For pure time refraction the condition $\omega_1 n_1 = \omega_2 n_2$ holds (where $\omega_1$ and $\omega_2$ are the angular frequencies before and after the time boundary and $n_1$ and $n_2$ are the indices before and after the time boundary).

This equation, in its differential form $d\omega/\omega = dn/n$, explicitly tells us that for a frequency conversion process to be relevant (i.e., experimentally measurable) a large change of the refractive index is needed on a very short time span. For this reason, despite the enormous potentials and intrinsic scientific curiosity around time-varying materials, consistent experimental advancements have been lagging. In fact, typical real-world temporal index changes are either fast (e.g. attosecond perturbations of bounded electrons) or large (e.g. material phase transitions) with these two attributes typically excluding each other [6].

Recently, transparent conducting oxides (TCOs) have been proven pivotal in the near infrared region to overcome the previously stated trade-off between amplitude and speed. Within this spectral window, these materials possess a bandwidth exceeding 260nm [7], [8] and allow for a 100% change of the refractive index triggered by single cycle pulses (6 fs) [8]. These remarkable results follow other previously reported key findings such as bandwidth-large frequency shifts [9], [10], [11], [12], unitary change refractive index [13], [14], dual colour hybrid nonlinearities [15], and many others [16], [17], [18], [19], thus creating a perfect playground for fully developing the potential of photon acceleration physics [20], [21], [22]. These materials also lend themselves towards the realisation of photonic time crystals, which can be theoretically exploited also for light amplification [23], [24], [25].

In this work, an ultra-thin time-varying layer is attained by optically pumping a sub-wavelength film of aluminium zinc oxide (AZO) operating in its near-zero-index (NZI) region. The optical excitation induces a large refractive index time gradient which is first ascendent and subsequently descendent thus exerting opposite effects on the front and back of a synchronised probe. Therefore, the transmitted pulse is spatially split into two halves each containing about half of the overall spectral power. In addition, these two "fission products" shows strong spectral shifts, which are opposite in sign for both halves and centred at the NZI carrier wavelength.

Our analytical model is an adaptation of the generalised spatio-temporal Snell's law [26] to subwavelength time varying layers, and it can quantitatively grasp all measure macroscopic parameters such as spectral shifts, angular shifts, and energy repartition. This spatio-spectral fission presented here is one of the most extreme nonlinear optical phenomena recordable on deeply subwavelength propagation lengths. These findings open new frontiers in the temporal engineering of material optical properties while providing insights into previously reported results that do not account for the spatial distribution of the propagating pulse.

# 2-Nonlinearities in time-varying systems

## 2.1- Time-varying systems vs classical nonlinearities

Although standard optical nonlinearities can lead to spectral energy redistribution, they can conceptually differ from nonlinear events occurring in time-varying media, as the latter are non-resonant, do not no submit to selection rules (i.e. phase matching), and exhibit a scattering cross section equal to 1 (i.e. it will affect all propagating photons) [3]. All these subtle differences can be intuitively understood by starting (as usual) from the harmonic oscillator.

Let us now consider the classical nonlinearities associated to bound electrons, which emerge from the anharmonic oscillations far from the equilibrium point. Consequently, only along a limited range of the electron motion nonlinear harmonics are generated. This oscillator can be modelled by the following Duffing equation [27]

$$\frac{d^2P}{dt^2} + \gamma\frac{dP}{dt} + \omega_0{}^2P + \beta(P\cdot P)P + \Theta(P\cdot P)^2P + \cdots = \omega_p{}^2E \tag{1}$$

Where E is the electric field, P is the polarization of the material, $\gamma$ is the scattering frequency, $\omega_0$ is the resonance frequency, $\omega_p$ is the plasma frequency, and $\beta$ and $\Theta$ are first and second order tensors describing the nonlinear restoring force, respectively. This mathematical representation immediately recalls into mind the standard polynomial expansion of material polarisation describing optical nonlinearities in bounded electrons [28].

On the other hand, hot electron nonlinearities stem from changes of the electrons effective mass [6], [29]. In fact, when pumping a metallic material, the electrons in the conduction band are heated up thus modifying their effective mass which in turn changes the plasma frequency, and ultimately modifies the refractive index. This is what happens in TCOs for a generous bandwidth around the cross-over wavelength, where associated macroscopic free electron nonlinearity can be described by:

$$\frac{d^2P}{dt^2} + \gamma\frac{dP}{dt} = \omega_p(t)^2E \tag{2}$$

In this second differential equation, $\omega_p$ is now a function of time through the transient and optically induced heating of free carriers (see Supplementary materials – Appendix A). In this second case, the nonlinearity is no longer directly dependent on the electric field and associated material polarisation, and the underlying process is always “active” thus leading to a unitary scattering cross section. For all these reasons, Eq. 2) implicitly describes the nonlinear optical behaviour of TCOs and justify why they are the best possible approximation to ideal time-varying systems [7][8] (See supplementary materials appendix A for further discussion).

In these compounds, we depart from a local view of the material response, concerned with electron-photon interaction, and consider light that is immersed in a uniform medium that changes its impedance with time. These differences between time refraction and more ”traditional” nonlinear optical processes, can lead to largely unexplored research territories extending in the domain of non-reciprocal systems [26], quantum optics [30] [31], time crystals [8], [24], [32], [33], and beyond [34] [35], [36], [37]. A great part of the quantitative analysis that follows in the present manuscript, starts from an experimentally retrieved time profile of the refractive index, which is induced by an intense pump trespassing a thin film of aluminium zinc oxide. The correspondent index time gradient will be directly linked to the observed spatial redistribution of spectral energy without any consideration on the microscopic material attributes leading to such extraordinary nonlinearities. For a deeper analysis of the nonlinear optical properties in TCOs we refer the reader to [38].

## 2.2-Pulse propagation through time-varying layers

Within the context of time varying materials, the dual case of an electromagnetic wave trespassing a spatial interface, is a radiation which experiences a sudden and global change of the index in time. While the former case pertains the well-known Snell’s law where energy is conserved (i.e., $\omega_1 = \omega_2$), the latter deals with the concept of time refraction, where instead momentum is the invariant parameter (i.e., $k_1 = k_2$).

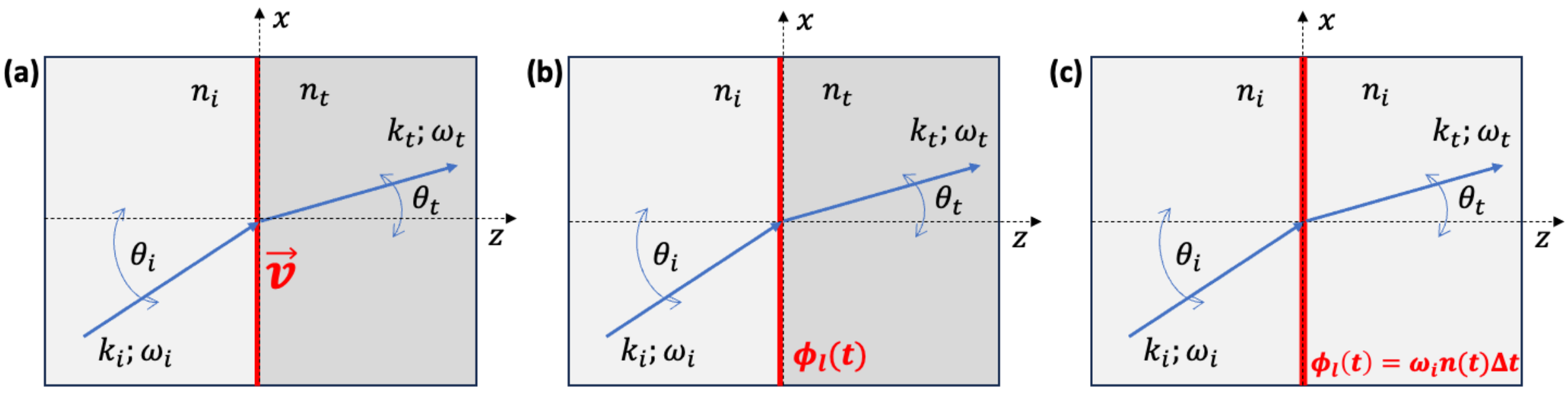

Fig. 1 - Spatio-temporal refraction and Snell's law adaptation. (a) Generalised case for an ideal merge between pure spatial and pure temporal refractions as described in [5]. (b) Intermediate model as represented in equation (8) which could adequately represent the case of a metallic metasurfaces on top of a nonlinear substrate. (c) Final adaptation of Snell's law for time-varying subwavelength thin layer which can adequately model the case of optically pumped TCO thin films..

A more general situation can be considered where the medium is both non-uniform in space and non-stationary in time. When this condition is achieved by combining ideal space and time refraction an equivalent model can be considered where the interface is moving at a given speed $\vec{v}$ normal to the interface (see Fig.1-a) [3], [39]. The Associated generalised Snell's law can be written as [26]:

$$n_1\omega_1 sin\theta_1 = n_2\omega_2 sin\theta_2 \tag{3}$$

Where $n_1$ and $\omega_1$ are the index and frequency before the moving boundary and $n_2$ and $\omega_2$ are the index and frequency after the moving boundary.

However, when the non-stationary system under analysis is a non-moving, and optically pumped, thin film of highly nonlinear low-index conductive oxide, the generalised and idealistic model proposed in [5] requires some adaptations. Indeed, the new model we seek should account for the deep subwavelength nature of the system in use. In fact, at the lowest value of the refractive index the operational effective wavelength in conductive oxides is several times the film thickness, which is set to be sub-micron also due to the material losses. Before adapting Eq. 3 to this more specific and realistic case, we can consider the intermediate scenario represented in Fig.1b. Here, a non-moving spatial interface, which is naturally defined by two materials with different static indices $n_i$ and $n_t$, induces an additional time-varying phase $\phi_l(t)$ term to the trespassing radiation. Such system is somehow difficult to find in the real world, however, a metallic metasurface patterned on top of a dielectric nonlinear substrate could fit the bill. For this device to operate as represented in Fig.1b, the nano-antennas should only act as field enhancer to localise the nonlinearities at the interface. An attempt to reproduce such system is reported in [40] where non-reciprocal behaviour is recorded for a beam reflected by an optically pumped nonlinear grating. However, the device used is limited in bandwidth (thus tested in a quasi-monochromatic regime), requires complex material fabrication processes, and need the combined effect of two beating pump waves to induce a relevant temporal phase change (maybe we add something to stress on how better we are?). A suitable theoretical discussion on equivalent system is reported in [26].

Let us now lay down the basic mathematical steps to deduce an adaptation of Snell's law for our specific case of a probe pulse trespassing a subwavelength thin film of a TCO layer whose refractive index varies in time under the effect of another ultra-fast pumping pulse. Since the film thickness is about 1/30$^{th}$ of the spatial extent of the pump pulse envelope, the maximum temporal index change is about 40 times bigger than the maximum spatial index change over the film thickness. For this reason, the medium can be considered uniform along the direction of propagation z.

In the geometric optics approximation, the electric field is written as $aEe^{i\phi}$, where $a$ is a slowly varying envelope, and $\phi$ is a spacetime dependent phase. The frequency and the wavenumber of the propagating beam are defined to be $\omega \equiv -\frac{\partial\phi}{\partial t}$ and $k \equiv \nabla\phi$, respectively.

For this system, a "phase conservation law" can be written at the interface which will look like:

$$\phi_t = \phi_i + \phi_l \tag{4}$$

where $\phi_i$ and $\phi_t$ are the incident and the transmitted phases of the probe, respectively and $\phi_l$ is the additional phase contribution induced by the time-varying layer. This phase conservation law generalises Snell's law and allows the simultaneous breaking of both spatial symmetry (due to the physical presence of the TCO layer) and temporal symmetry (due to the time-dependent refractive index that the pump induces on the TCO layer). With obvious formalism, we can now specialise our formulas to both energy and momentum conservation, by taking time and space derivatives of Eq. (4):

$$\omega_t = \omega_i - \frac{\partial\phi_l}{\partial t} \tag{5}$$

$$k_{t,x} = k_{i,x} + \frac{\partial\phi_l}{\partial x} \tag{6}$$

Under the assumption that the spatial index change along the interface is negligible (in experimental terms this correspond to a pump pulse which is considerably larger than the probe), we simply have $\frac{\partial\phi_l}{\partial x} = 0$. It is worth noticing that this term is non-zero for the case of metasurfaces, which could lead to a further generalisation of the modified Snell's law reported in Eq. (3) [26]. However, in absence of spatial phase gradient along x (i.e., no metasurfaces), Eq. (6) gives:

$$k_t sin\theta_t = k_i sin\theta_i \tag{7}$$

where $k_i$ and $k_t$ are the total wavenumbers of the incident and the transmitted waves, respectively. Now the crucial point is the expressions of $k_i \equiv \omega_i n_i/c$ and $k_t \equiv \omega_t n_t/c$ which lead to:

$$\frac{n_t}{c}\left(\omega_i - \frac{\partial\phi_l}{\partial t}\right) sin\theta_t = \frac{n_i}{c}\omega_i sin\theta_i \tag{8}$$

At this point, equation (8) is well suited to represent the case depicted in Fig.1-b, which as previously discussed could well reproduce the optical behaviour of a plasmonic metasurface patterned on top of a nonlinear substrate. To complete the model adaptation to the case of an air/TCO optically pumped interface we now need to add the assumption that $n_i = n_t$ since the TCO layer is deeply subwavelength (i.e., film thickness < effective wavelength), thus obtaining:

$$\left(\omega_i - \frac{\partial\phi_l}{\partial t}\right) sin\theta_t = \omega_i sin\theta_i \tag{9}$$

from which one derives the angle of transmission:

$$\theta_t = arcsin\left[\frac{sin\theta_i}{1 - \frac{1}{\omega_i}\frac{\partial\phi_l}{\partial t}}\right] \tag{10}$$

If there is no time variation of the refractive index (e.g., switching off the pump), the contribution of the TCO layer to the phase would vanish, ($\frac{\partial\phi_l}{\partial t} = 0$) leaving $\theta_i = \theta_t$, as expected. These two final equations can be associated to the scheme in Fig.1c, where a subwavelength time-varying layer (representing our AZO thin film) is immersed in an environment with homogenous and stationary index.

Now an important step is specifying the phase $\phi_l$ accumulated by the incident wave during the interaction of the probe with the TCO layer. One natural guess is $\phi_l(t) = \omega_i n(t)\Delta t$ where n(t) is the time variation of the refractive index in the medium due to the pump and Δt is the interaction time of the probe and the layer. All this considered, the transmitted angle can be rewritten as a function of the index time gradient as:

$$\theta_t(t) = arcsin\left[\frac{sin\theta_i}{1 - \frac{\partial n(t)}{\partial t}\Delta t}\right] \tag{11}$$

Let us now consider what happens to the wavelength of the incident radiation trespassing our time-varying layer. The transmitted wavelength can be written as a function of the index time gradient by taking the derivative of $\phi_l(t) = \omega_i n(t)\Delta t$ and substituting it into Eq. (5):

$$\omega_t(t) = \omega_i\left(1 - \frac{\partial n(t)}{\partial t}\Delta t\right) \tag{12}$$

These last two equations will be recalled along the manuscript to describe and explain all reported results pertaining the spectral and spatial redistribution of energy for an ultra-fast pulse crossing a time-varying layer.

## 3-Experimental set-up and data processing

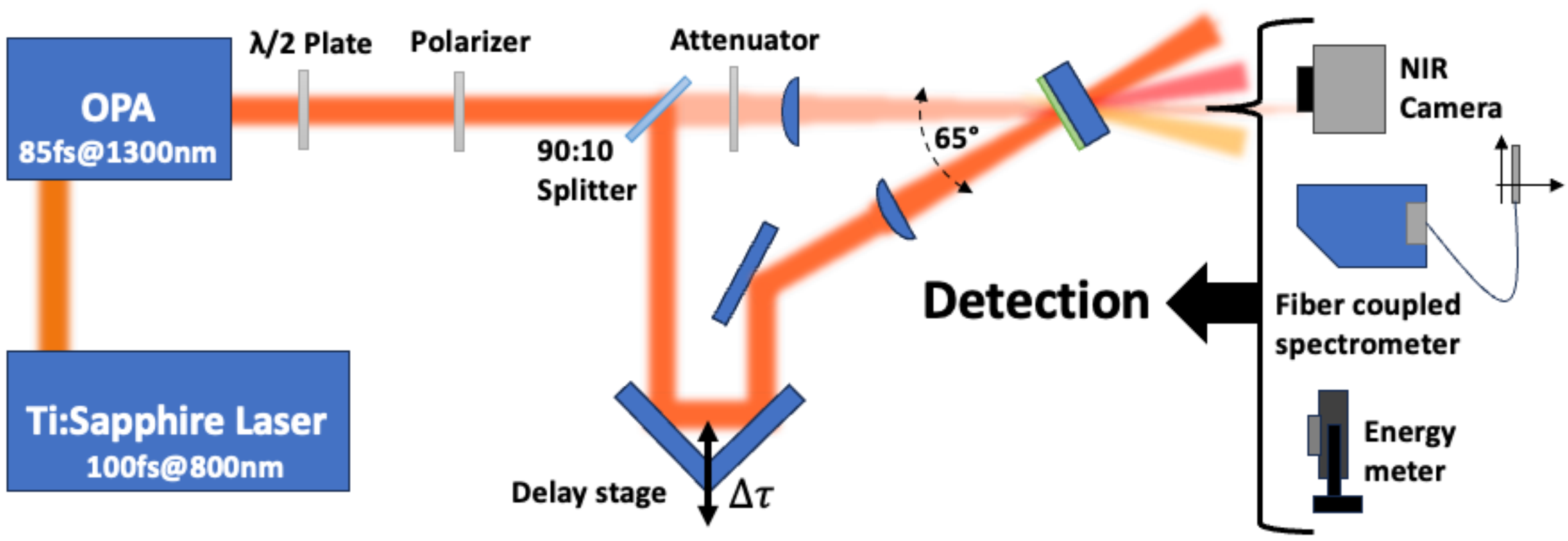

Fig. 2 – Experimental set-up. Pump & probe apparatus operating in a degenerate configuration (i.e., both pump and probe photons have same photon energies). The pump pulse has the role of inducing a strong and ultra-fast temporal change of the refractive index and turning our oxide film into a time-varying subwavelength layer. The schematics also shows spatio-spectral fission of the probe pulse under opportune synchronisation between probe and material response.

Experiments have been conducted using a pump/probe setup (See Fig. 2) employing 93 fs pulses (See supplementary materials appendix B for FROG characterisation) at a repetition rate of 1 kHz and centred at 1300 nm, which corresponds to the epsilon-near-zero (ENZ) crossover wavelength of our 900 nm thick low-index (n = 0.35) AZO film (See [41] for characterisation). The choice of the operational wavelength close to the crossover point is well justified by the relevant literature reporting on enhanced nonlinearities in ENZ media. The same applies to the decision of using a degenerate configuration for our experiments [42], [43]. The pump beam is focused onto the sample at normal incidence with peak power of about 1.8 $TW/cm^2$ (well below the damage threshold [38], [41]). The probe beam is attenuated well below the threshold for self-action nonlinearities and focused onto the sample at an angle of 65°. The condition of a steep angle of incidence for the probe is chosen for various reasons: i) to maximise local field enhancement [44], [45] ii) to size the material time response with respect to the probe duration for optimal spectral/spatial fission; iii) to maximize the relative induced change in transmission, thus helping to detect spatio-spectral variations in the refracted signal; iv) to provide the strongest angular shift according to eq.(11). The value of 65° was chose via experimental means by looking at the angle maximising nonlinear beam deflection. The number is also consistent with similar values reported in literature to maximise nonlinear effects in different TCOs [13], [45].

We should stress that the angle of incidence is very critical to achieve full spatio-spectral fission because it alters the relative time duration of probe with respect to the material response (and Pump duration). For this reason, there is an optimal angle that optimise the temporal overlap between the material response (n(t)) and the probe temporal profile (see Fig.4 central panel third row). In this way, we ensure to maximise the fraction of the probe energy affected by the material nonlinearities (dn/dt). However, this condition has to be balanced with another issue. In fact, if the incident angle is too large, we quench transmission and the correspondent signal to noise ratio; but if it is too narrow, then we reduce the nonlinear angular separation (see Eq.10 and supplementary material appendix C). Given the critical balance among all these factors, the optimal incidence is evaluated experimentally by testing which angle produces the strongest spatio-spectral separation with optimal signal to noise ratio

Information about the refracted beams have been collected simultaneously using different tools such as a high resolution infra-red camera, fiber coupled spectrometers, and energy meters for multiple values of the pump-to-probe delays $\Delta\tau$. The spectrometer was mounted on a calibrated motorised stage which allowed for spatial scan of the spectral distribution at given x values along a horizontal axis orthogonal to the probe propagation direction.

From equations (11) and (12) we see that as the probe pulse trespasses the time-varying layer, both its spectrum and shape must be affected by the index time gradient. To provide a global picture of what happens to both energy and spectrum of the transmitted probe, the information acquired by both camera and spectrometer were combined. This unification was attained by assigning to each x point one unique wavelength evaluated as the centre of mass of the correspondent spectrum. This analytical process is depicted for a given $\Delta\tau$ in Fig.3 where, in the lower panel, the spatial probe profile is mapped to its local spectral distribution (see details in Methodology). Here, the x-axis is converted into units of angular deflection using $\Delta\theta = \tan^{-1}\frac{x}{L} - \theta_i$, where $\theta_i$ is the incidence angle and L is the distance from the sample to the camera, which we set to 7.5 cm. This data representation will be used in the next section to reconstruct a full transient spatio-spectral picture of the refracted probe pulses.

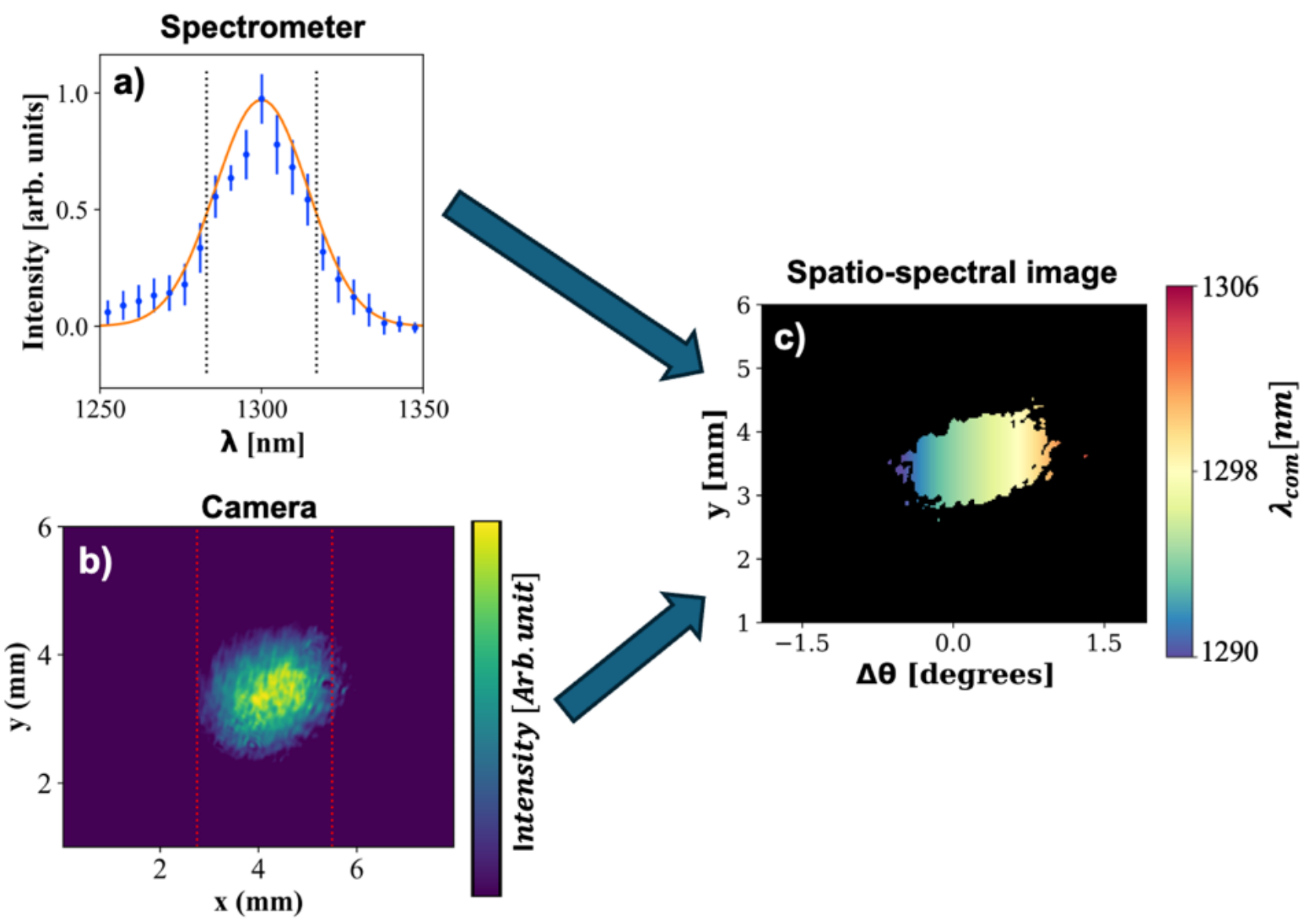

Fig. 3 – Process for creating spatio-spectral images from spectrometer measurements and camera images. (a) Top left panel provides an example spectrum of the transmitted probe pulse for a given $\Delta\tau$ time delay and a given horizontal spatial position the (b) bottom left panel gives the spatial intensity profile as measured by a near-infrared camera. The (c) right panel shows the full spatio-spectral profile of our probe pulse at a single pump-probe delay. Reported plots and images are taken for $\Delta\tau = -206\ fs$).

# 4-Results and Discussion

## 4.1-Spatio-spectral fission

Following the previously outlined data analysis, we can plot the simultaneous spatial and temporal evolution of both energy and spectrum for a 93 fs quasi-transform-limited pulse trespassing an optically pumped subwavelength layer of AZO. This study is reported in Fig.4.where four rows of panels show what happens to a pulse as it goes through a time varying layer exhibiting two consecutive and abrupt changes of refractive index, which are also opposite in sign. The outcome is indeed determined by the specific pump-to-probe delay Δτ. The first row of panels shows numerical predictions about the probe evolution trespassing the AZO layer for different pump-to-probe time delays. The second row of panels reports about the correspondent experimentally acquired results. For the numerical evaluation, the probe beam is approximated with a spatially discretised k-vector distribution associated to a focused, spatially unchirped, and transform limited gaussian pulse centred at 1300 nm. This distribution is then used as input for equations 11 and 12 where the refractive index temporal profile has been evaluated as reported in the subsequent section "Energy redistribution" of the present manuscript.

From the first two rows of panels in Fig.4, at large values of time delay ($\Delta\tau = -206fs, +279fs$), the probe does not interact with the refractive index perturbation imparted by the pump, and thus there is no shift (neither spectral nor spatial). Only a small wavelength deviation across the spatial width of the pulse is noticeable, which can be attributed to a small spatial chirp originating in the optical parametric amplifier (OPA). In this case the pulse propagation direction is unaltered by the unpumped AZO layer. The quasi-transform-limited nature of both incident optical pulses is also proved by accurate FROG measurements (See Supplementary materials – Appendix B)

If the pulse is slightly late ($\Delta\tau = -30fs$) with respect to the material response ($dn\ /\ dt$), its front sees a positive gradient as schematically depicted in the first panel of third row in Fig.4. The transmitted pulse is then red-shifted and re-routed to the right with respect to its original incidence direction (second panels of first two rows in Fig.4). The dual case occurs if the incident pulse is mildly early ($\Delta\tau = +61fs$) compared to the material response (third panel of third row in Fig.4). In this case, it experiences a negative index time gradient that deflect the pulse on the left while blueshifting its spectrum (fourth panel of first two rows in Fig.4). The most extreme situation though pertains the case when pump and probe are optimally overlapped in both time and space ($\Delta\tau = 0$). In this case, the probe is literally torn apart as its front and back are pulled in opposite direction by opposite index gradients. The two resulting halves are detoured away from the incidence direction while being red- and blue-shifted, respectively (central panels of first, second, and third row in Fig.4 and fourth row of panels).

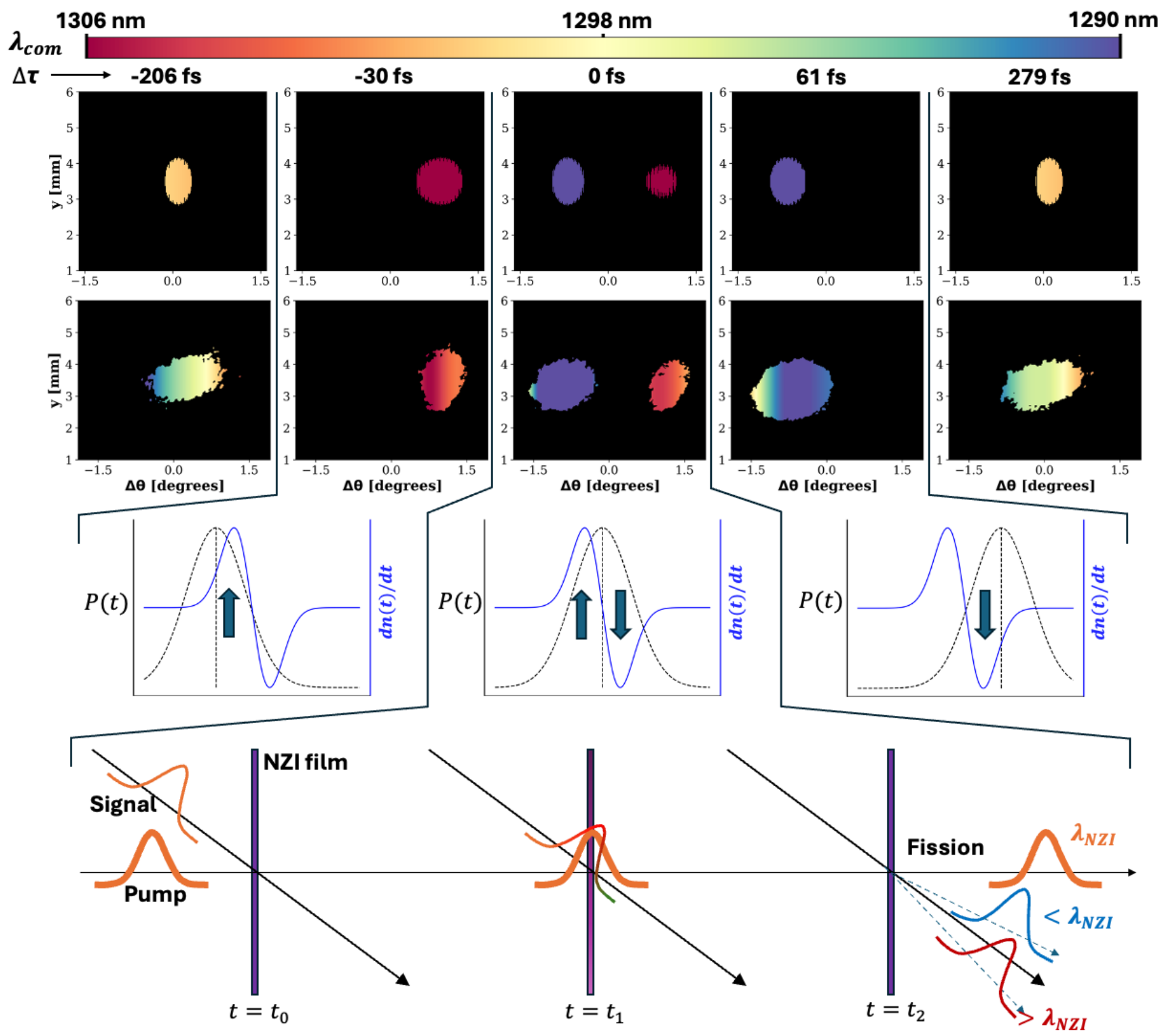

Fig. 4 – Full spatio-spectral information of transmitted probe pulse as the pump-probe delay is tuned. Panels report about the probe transmitted beam profiles (with the x-axis recalibrated in deflection angle Δθ) together with the associated spatial spectral distribution for different values of the pump-to-probe time delay (Δτ). Given the problem symmetry with respect to the incidence plane, only a horizontal scan of the spectral distribution is performed and one single wavelength value (i.e., the centre of mass of the associated spectrum) is assigned for every given x (Δθ), which is the same value for all associated y. The middle row of panels shows black (dashed) and blue (solid) curves representing the probe temporal profile and the material response (dn⁄dt) for specific pump-to-probe delays. Finally, at the bottom, a three-panel scheme represents, in a simplified manner, the full spatio-spectral fission of probe pulse when Δτ=0, which is when probe and the pump peak are synchronised. For Δτ=-30fs the leading edge of the probe experiences a positive refractive index time gradient, which causes the probe to redshift by about 4 nm and deflect on the right by +1.15 degrees with respect to the direction of the incident probe. For Δτ=0 the maximum pump-probe overlap is achieved and the probe splits into two halves with a +0.85 and -0.7 angular deflections, respectively. These two beams are characterised by different spectra corresponding to blue-shifting and red-shifting the two halves of the original spectrum with respect to its central wavelength. For Δτ=+61 fs the back of the beam experiences a negative index gradient, and it is then detoured away from the original incident direction with a maximal angular deflection of -0.87° and a centre of mass blue shifted of about -7 nm. Finally, for Δτ=279 fs the two pulses are no longer overlapped in time and the transmitted probe is back to the unpumped case.

It should be mentioned that this splitting is not perfect as there is a <10% residual optical power in between the redshifted and blue shifted beam. The origin of this "leakage" stems from the simple observation that as the index of the time-varying film undergoes an up and down transition, the material will assume a time-invariant form for a brief instant when $dn / dt = 0$, thus leaving some part of the probe beam unaltered according to eq. (11) and (12). This residual power does not appear in our figure being below the set noise threshold.

Regarding the reported results, it is worth underlying how remarkable is the agreement between theory and experiments despite the simplicity of the model, which does not account for dispersion and losses thus neglecting the effects that the nonlinear change in the imaginary index could have on both spatial and spectral redistribution. However, previously reported studies, looking into the complex nonlinearities of AZO films for very similar excitation conditions, show that although an increase of the real index is accompanied by a contextual reduction of the associated imaginary part (as also dictated by Kramers-Kronig relations), the magnitude of the former is considerably larger that the latter thus making our simplified analysis robust against experimental tests.

## 4.2-Energy redistribution

One limitation of the spatio-spectral analysis reported in Fig.4 is that it shadows the acquired information about the energy redistribution mediated by the time-varying medium. To address this, we analyze the probe transmitted power repartition in both spectrum and space against the pump-to-probe delay. This repartition is reported in Fig.5 where both acquired spectra and camera images are partitioned into three distinctive regions which allow for appreciating the redistribution of energy as the probe is affected by the time-varying layer. On the one hand, the power spectral density separation can be described by plotting three curves (Fig.5-a): i) the percentage of transmitted power below 1285nm (blue crosses); ii) between 1285 nm and 1315 nm (green circles); iii) and above 1315 nm (red dots). On the other hand, the spatial power separation is considered in Fig.5-b where the three plots represent: I) the percentage of transmitted power below -1° of deflection (blue crosses); between -1° and +1° (green circles); and III) above +1° (red dots). Finally, by acquiring information about the total percentage of transmitted power, the plot in Fig.5-c is attained, which shows an increase in transmission over 300%.

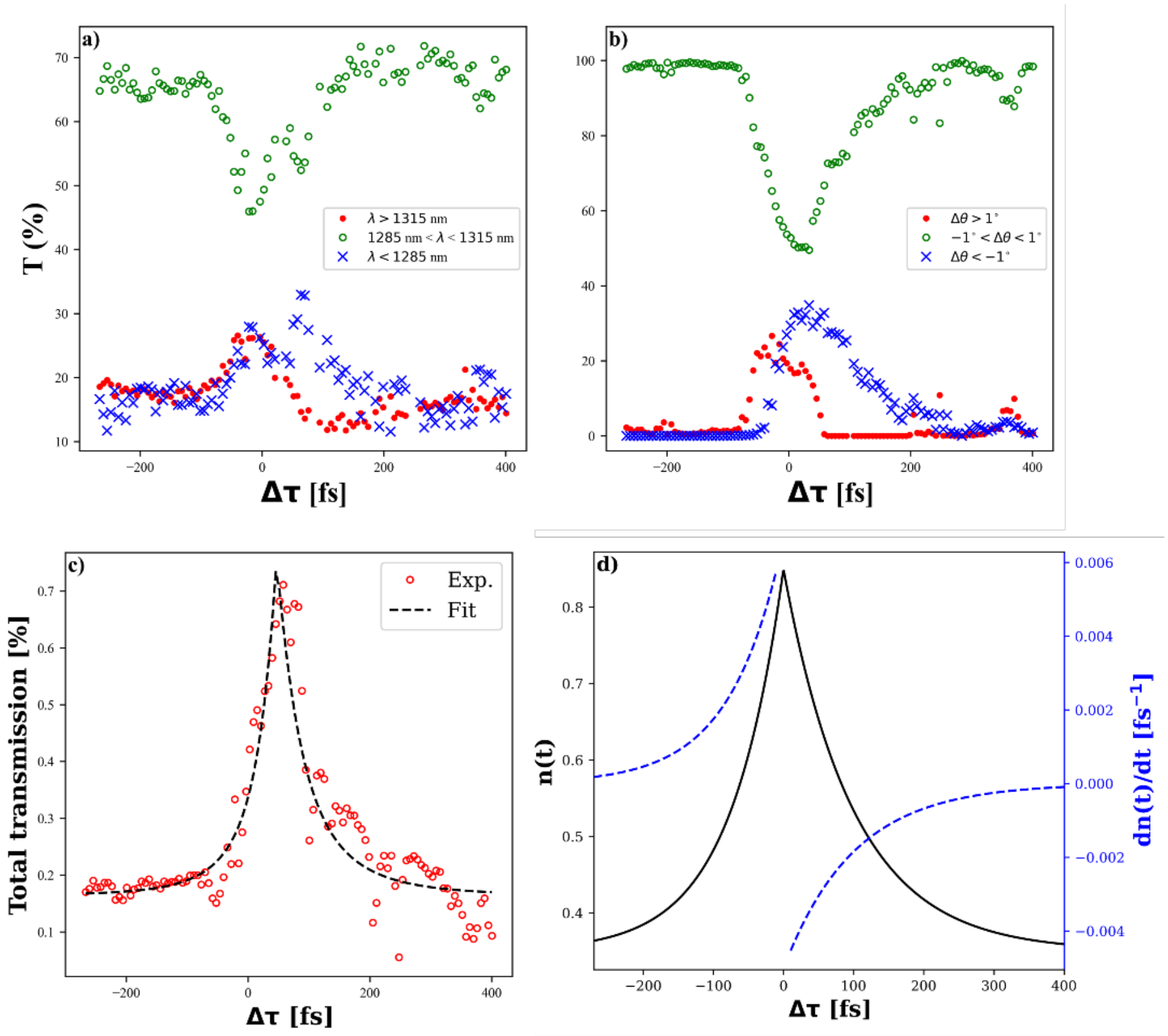

Fig. 5 – Probe transmitted power repartition in both spectrum and space against the pump-to-probe delay. a) spectral partitioning of transmitted power b) spatial partitioning of transmitted power. In both panels blue crosses correspond to blueshifted and deflected away from the film's normal while the red dots are for redshifted and deflected towards the film's normal. Black stars provide the total transmitted power T, normalised to 100% for transmission in the linear regime.

This apparent increase in transmitted power is due to a reduction in the reflection at the interface under optical excitation, which leads to a temporary reduction in the air/AZO index contrast while also helping highlighting the energy splitting through the time-varying layer. In other words, by drastically increasing the index we are not only creating a time varying medium but also moving the refractive index much closer to that of air, dramatically improving transmission through the film and increasing the amount of time-refracted light.

The nonlinear index modulation shown in Fig. 5d was recovered from the transient transmittance shown in Fig. 5c. First, the relative transmission data is converted in absolute values of transmission using previously reported linear transmission measurements for this exact film [41]. A gaussian rising index (driven by the temporal profile of the gaussian pump), and an exponentially falling index (due to the relaxation of hot electron nonlinearities) are fitted to the transmission data (a transfer matrix approach is used to convert the fitted indices into transmission values). The extremely low absolute transmission of the probe is due to the high angle used in our experiments and the large index differential between air and AZO. Imaginary index is held constant at the linear value in this analysis.

## 4.3-Angular deflection

The normalised angular intensity distribution of the probe as a function of the pump-to-probe delay $\Delta\tau$ is shown in colourmap in Fig. 6-a. When reading this figure from left to right (increasing values of $\Delta\tau$), we initially operate in the linear regime with no overlap between pump and probe and no observable angular deflection. As $\Delta\tau$ progresses, pump and probe start overlapping in time, and we enter the redshift dominant phase which is accompanied with a positive angular deflection. Next, at a pump-to-probe offset close to zero, a clear spatial splitting regime appears where in time half probe (front of the pulse) overlaps with a positive index gradient and the other half (back of the pulse) with a negative gradient. This regime stays dominant for about 45 fs (measured as the duration within which the absolute intensity difference between the right and left deflected beams is < $1/e^2$).

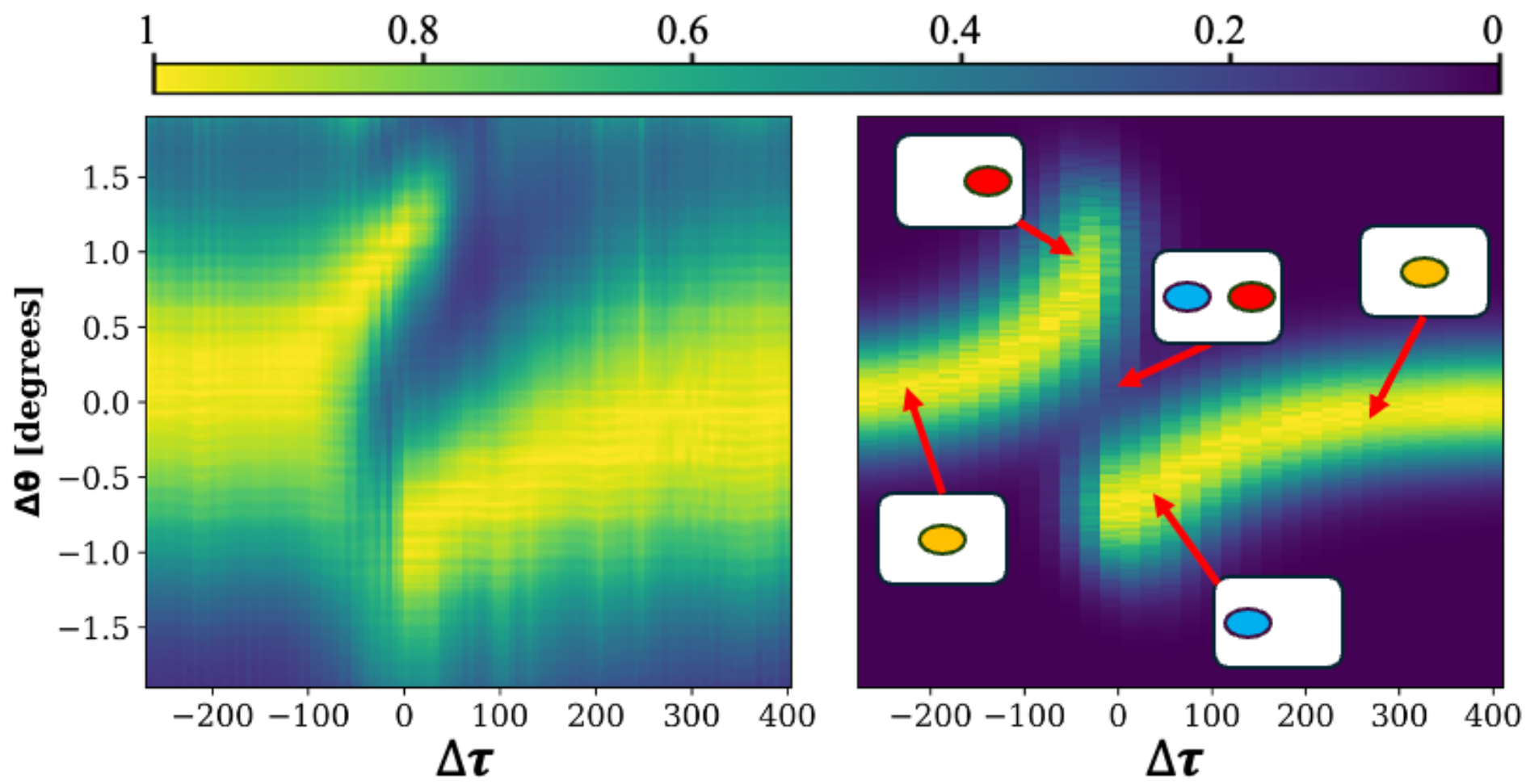

Fig. 6 – Power distribution as a function of deflection angle and pump-to-probe delay. a) Heatmap of the angular power distribution as the pump-probe delay is tuned. the picture displays the normalised heatmap of intensity vs pump-probe delay and angular shift and it is attained by integrating the camera pixel intensity $\boldsymbol{I_c}(\Delta\tau, \Delta\theta,\ y)$ along the y-axis. The set of five little icons at the top of the picture mark the temporal delays corresponding to the spatio-spectral plots reported in Fig.4. The blue line overlayed on the heatmap is found using Eq. 12, the refractive index modulation shown in the right panel, and the same numerical procedure adopted for the theorical panels in Fig.4. b) Refractive index modulation as recovered from transmission curve in Fig. 5. The asymmetrical nature of AZOs nonlinear index response can be clearly seen as the redshift (rising fit) is faster and stronger than the blueshift (relaxation fit).

Finally, when the probe only interacts with the negative index time gradient, a pure blueshift (negative angular deflection) is clearly observable, which is longer in duration than the redshift phase. This is due to the asymmetric rise and fall time of the hot electron nonlinearities we are invoking.

Using previously acquired nonlinear characterisation data from our AZO film coupled with transfer matrix formalism and the total transmission curve (black star markers, Fig. 5), we recover the refractive index modulation induced by the pump. The index profile recovered from this analysis (Fig. 6-b) is composed of two separate exponential fits (Excitation fit – blue solid curve; Relaxation fit – red dashed line) with different time constants. The index's rising time is about 75 fs while the relaxation time is 100 fs, which match closely with those reported in the literature [41]. When inputting the recovered index profile in Eq. 12, and accounting for the pump-to-probe spatial/temporal overlap (as previously discussed with regards to the theoretical panels in Fig.4) a $\Delta\theta(\Delta\tau)$ function is attained (blue solid line overlayed onto the heatmap in Fig.6). As we can see, the predicted angular deflection closely agrees with the underlying experimental data (note that a $\Delta t$ of 3 fs is used in Eq.12, which is consistent with the film thickness).

From the retrieved refractive index profile, we can also estimate the maximum induced wavelength shifts to be 11.26 nm redshift and 6.97 nm blueshift, which is in good agreement with previously reported studies in the literature [10]. However, when using our centre of mass metric for high incident angles, spectral shift becomes smaller than those directly predicted by our simplified model applied to a single peak wavelength (see further details in Supplementary material Appendix C). Additionally, it is worth considering absorption, which disproportionately attenuates redshifted light.

Regarding the results reported in the present manuscript, it is important to highlight that, although significant efforts have been made to investigate the nonlinear modulation of spectrum and directionality of a propagating beam, this is the first time these two effects have been observed simultaneously on a macroscopic scale with comprehensive beam reshaping distribution information. Besides the intrinsic significance of our combined spatio-spectral analysis, which opens new avenues for the temporal engineering of material properties, our work provides a deeper understanding of key findings recently reported in the literature. For instance, in [46] although both spatial and spectral analyses are performed for a probe beam trespassing a time-varying layer (i.e., optically pumped ITO), these were carried out by either integrating over space (i.e., collecting all the transmitted light and sending it into a spectrometer), or by integrating over the spectrum (i.e., by measuring optical power via colour-bling photodiodes), an approach which might miss the overall complex spatio-spectral reconfigurability of the probe. To clarify this last point, we can underline to the reader that if instead of performing a detailed spatially-resolved spectral analysis of the probe we had simply sent all transmitted light into a spectrometer, the spatio-spectral fission reported in Fig.4 (central panels of first and second rows), would have been misinterpreted as a simple spectral broadening. Another key factor arising from comparing our work with [46] and other works on the subject [???], is that the fabrication conditions for maximising the time-varying attributes of TCO films, can be very critical. For instance, to attain spatio-spectral fission a material that exhibits both ultra-fast excitation and recombination times is required.

From equation 11) and 12) is clear that our capability to temporal-engineer the material for optical beam shaping and spectral re-distribution is directly link to the refractive index time gradient. This parameter in TCO is the highest possible with experimentally proven effective refractive index change of about 0.5 in only a few fs (few cycles pulse) [8]. In addition to this, due to the hybrid nature of TCOs, the material response time can be forcefully shortened by the combined use of interband and intraband optical pumping [15]. Thanks to the unprecedented nonlinear optical properties of low-index conducting oxides we have now the freedom to transfer our capability in engineering the optical excitation (duration, shape, chirp, etc. of the pump) into the possibility to desig a specific time-varying material.

# 5-Conclusions

In conclusion, we report on splitting of an ultra-fast 93 fs pulse centred at telecom wavelength into two almost-equal parts as it trespasses a submicron-thin time-varying layer of transparent conducting oxide. This sudden, and almost complete, split shows a significant angular separation >2° (1.15° for redshifted beam and -0.87° for blueshifted beam) which is accompanied by a strong spectral fission and a large (> 300%) transmission enhancement. A simplified model has also been developed that can quantitatively grasp both the spectral and spatial time refraction with the only fitting parameters being the time refractive index profile induced by the pump, the group index, and the interaction time $\Delta t$. From this analysis, a nonlinear index profile was recovered that closely matches those measured in previous reports and accurately fits our experimental data. These results provide fundamental insights about the remarkable potentials of these materials for application in all-optical integrated photonics, and for the further exploration of photonic time crystals, on chip nonclassical light generation, and integrated neural networks.

# 6-Methodology

## 6.1-Experimental approach

Pump probe setup used 93 fs pulses at a repetition rate of 1 kHz. Operational parametric amplifier (OPA) output was tuned to 1300 nm which is correspondent to the ENZ point of our 900 nm AZO film. The choice of operating close to the NZI The output is then reshaped using a telescope and is horizontally polarised. The beam is split in an 80:20 ratio, and smaller part is used as the probe arm. The pump beam is focused onto the sample at normal incidence with a 10 cm lens resulting in a 60 μm ($1/e^2$) spot size with peak power of 1.8 $TW/cm^2$. The probe beam is then attenuated to a negligible power level to exclude self-action nonlinearities and is then focused onto the sample at an angle of 65° with a 7.5 cm lens, producing a 30 μm ($1/e^2$) spot size on the AZO. Finally, the probe is measured with both a near-IR camera and a fiber coupled spectrometer at a focal length from the sample. The AZO film used in our experiment is 900 nm thick and was deposited onto the silica substrate in a low oxygen environment by pulsed laser deposition. Characterisation of the samples refractive index can be found in the left panel of Fig. 7. Further fabrication details can be found in [47]. Preliminary pump/probe tests performed on the bare substrate were used to clean up the reported results from any spurious nonlinear contribution eventually coming from the substrate. In this regard it is worth mentioning that despite the high pump intensity any effect on the probe from the base fused silica substrate was negligible. This should not be a surprise given that the nonlinear Kerr coefficient in AZO is over four orders of magnitude higher the one in silica. In addition to this, the signal trespassing the 900 nm of AZO arrives considerably attenuated at the AZO/Fused-silica-Substrate interface, thus further limiting any nonlinear contribution from the substrate and making our analysis even more robust.

## 6.2-Numerical verification

The aim of this simulation was to verify the infinitely thin film approximation we use in our simplified model. Numerical verification experiments were completed via a second order FDTD scheme for Maxwells equations coupled to an ADE method for modelling the AZO's dispersion [48]. A nonlinear perturbation to the material was simulated by modifying the plasma frequency and damping constant of the Drude model at each time step in accordance with a gaussian index perturbation. The values for these parameterised changes were recovered from fitting linear Drude models to the pumped dispersions found in [41]. Finally, we interpolated this dispersion over a gaussian pulse of 93 fs that propagates through our material at the group velocity of our dispersion. The result of this comparison is shown in the right panel of Fig. 7, where both the FDTD simulation and simplified model deliver similar results. Additionally, in order to validate the existence of spatio-spectral fission when losses and dispersion are considered, we have applied a more comprehensive material model, described in supplementary material appendix D.

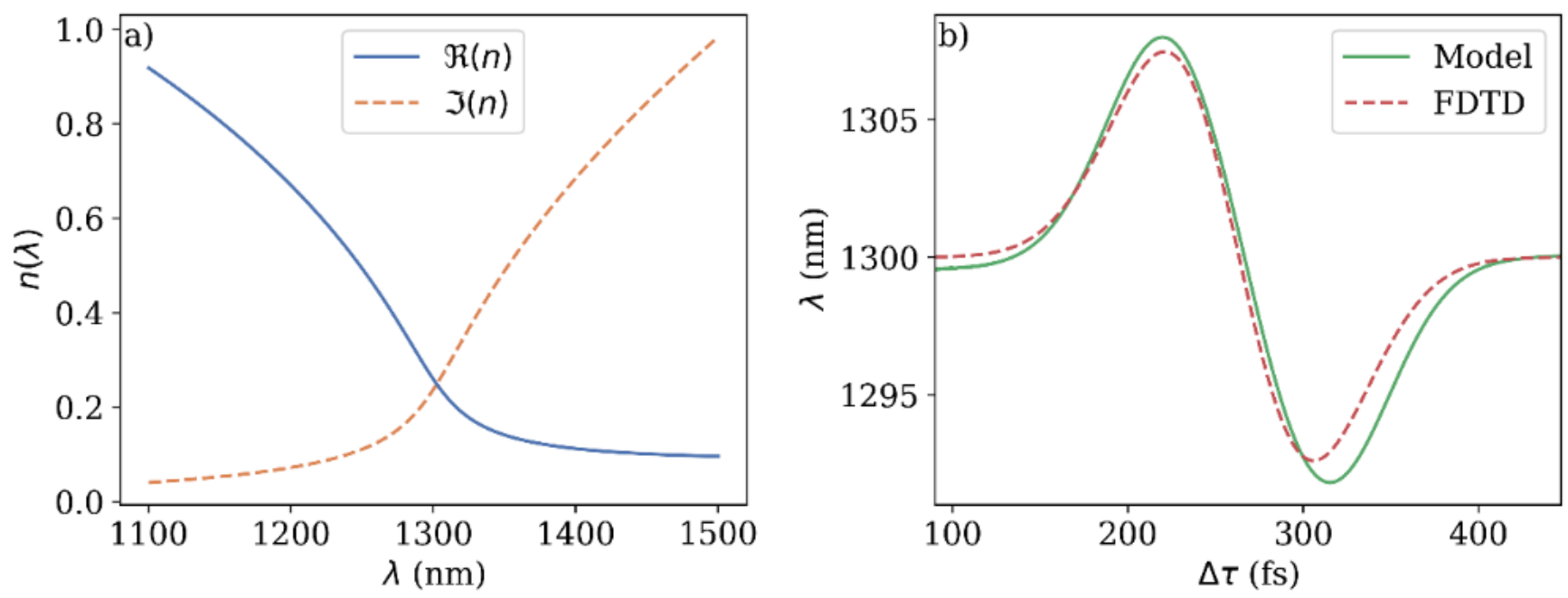

Fig. 7 – a) Linear refractive index of the 900 nm AZO sample used in experiments. b) Wavelength shift predicted from a gaussian index perturbation using the simplified model vs a FDTD approach.

## Author Acknowledgements

The Heriot–Watt team wishes to acknowledge economic support from EPSRC project ID: EP/X035158/1, and AFOSR (EOARD) under Award No.FA8655-23-1-7254. A.B. and V.M.S. acknowledge support by the U.S. Department of Energy, Office of Basic Energy Sciences, Division of Materials Sciences and Engineering under Award DE-SC0017717. All authors wish to thank Prof. Maria Antonietta Vincenti for useful discussions.